\documentclass[10pt]{article}
\usepackage{xcolor}
\usepackage{listings}
\usepackage{amsmath}
\usepackage{amssymb}
\usepackage{graphicx}
\usepackage{geometry}
\usepackage{verbatim}

\usepackage{mdframed}

\title{Optimal Linear Signal: An Unsupervised Machine Learning Framework to Optimize PnL with Linear Signals.}
\author{Pierre RENUCCI}
\date{Oct.-Nov. 2023}

\begin{document}

\maketitle
\begin{mdframed}
\begin{center}
    \bfseries \large Abstract
\end{center}
\smallbreak 
This study presents an unsupervised machine learning approach for optimizing Profit and Loss (PnL) in quantitative finance. Our algorithm, akin to an unsupervised variant of linear regression, maximizes the Sharpe Ratio of PnL generated from signals constructed linearly from exogenous variables. 

The methodology employs a linear relationship between exogenous variables and the trading signal, with the objective of maximizing the Sharpe Ratio through parameter optimization. Empirical application on an ETF representing U.S. Treasury bonds demonstrates the model's effectiveness, supported by regularization techniques to mitigate overfitting. The study concludes with potential avenues for further development, including generalized time steps and enhanced corrective terms.

\end{mdframed}\noindent The code of the model and the empiric strategy are available on my GitHub: \textsf{Cnernc/OptimalLinearSignal}

\vfill

\section*{Introduction}

In the field of quantitative finance, the creation of signals to generate Profit and Loss (PnL) is a key component. Our initial objective was to find a simple unsupervised machine learning (ML) algorithm capable of exploiting a set of exogenous variables to produce a PnL-effective signal. We observed that the literature in quantitative finance largely favors supervised ML approaches, aiming to predict specific financial magnitudes, such as asset prices, as highlighted in the works of Johnson (2023) \cite{Johnson2023}, Rouf et al. \cite{Rouf202x}, Soni et al. (2022) \cite{Soni2022}, and Kumar et al. (2022) \cite{Kumar2022}.

There are also unsupervised techniques, as presented by Kelly and Xiu (2023) \cite{KellyXiu2023}, and Hoang and Wiegratz (2023) \cite{HoangWiegratz2023}. However, these methods are very specific and lack the versatility to be used as general tools, specifically for small dataset. An exception in this category is the generation of signals via Principal Component Analysis (PCA), as explained by Ghorbani and Chong (2020) \cite{Ghorbani2020}. Nevertheless, this technique offers little flexibility and does not allow for the explicit optimization of specific criteria within the PnL.

We therefore turned our attention to the literature on optimization in finance, with studies such as those by Jurczenko et al. (2019) \cite{Jurczenko2019}, Huang et al. (2019) \cite{Huang2019}, and Cornuéjols et al. (2017, 2018) \cite{Cornuejols2017} \cite{Cornuejols2018}, Reppen et al. (2022) \cite{Reppen2022}, primarily focusing on portfolio optimization. However, these works are more concerned with optimizing portfolios than signal creation.

As a result, we decided to construct our own algorithm, developing an unsupervised counterpart to linear regression aimed at optimizing the Sharpe Ratio of a PnL. The model is based on two hypotheses of linearity: the linearity of the relationship between the exogenous variables and the signal; and the linearity of the relationship between the PnL and the signal. Using these, we can deduce a parametric representation of the PnL. We sought to optimize these parameters to maximize the Sharpe Ratio of the obtained PnL.

The optimal parameters are calculated over a certain training period and are then used to generate the signal subsequently. Other techniques, mainly regularization and correction of this signal, are then addressed and ultimately enable the creation of a very effective strategy on a backtest of about twenty years.

\newpage
\vfill
\begin{mdframed}
\begin{center}
    \bfseries \Large Executive Summary
\end{center}

\begin{itemize}
    \item We consider an asset with a specific price series $\textsf{price}_t$ and a set of \emph{stationary} and \emph{homoscedastic} variables $X_{1,t},\: X_{2,t},\: \ldots,\: X_{n,t}$. 

    \item The objective is to derive the 'optimal' linear signal extracted from this variable. We work under two linear assumptions: 
    \begin{itemize}
        \item We will consider the signal as a linear combination of the exogenous variables: 
        $$\textsf{signal}_t = \alpha_0 + \alpha_1 X_{1,t} + \alpha_2 X_{2,t} + \ldots + \alpha_n X_{n,t}$$

        \item At each time step $t$, the positions taken is proportional to both the signal and the price $\textsf{pos}_t = \textsf{price}_t \times \textsf{signal}_t$ (a negative value would correspond to a short position). This approach establishes a linear relationship between Profit and Loss (PnL) and the trading signal. We get, after computation: 
        $$\textsf{PnL}_t = \textsf{signal}_ {t-1} \times (\textsf{price}_t - \textsf{price}_{t-1})$$
    \end{itemize} 

    This allows us to express PnL as a linear combination of the variables: 
    $$\textsf{PnL}_t = \alpha^\text{T}[X_{t-1} \times (\textsf{price}_t - \textsf{price}_{t-1})]$$

    \item A parametric expression of the empiric Sharpe Ratio is then derived: 
    $$\mathcal{L}(\alpha) := \frac{\overline{PnL}}{\sigma(PnL)} = \frac{\alpha^\text{T} \mu}{\sqrt{\alpha^\text{T} \Sigma \alpha}}$$ 
    With $\mu_i$ and $\Sigma_{i,j}$ the empiric mean and covariance of the variables $(X_{i,t-1} \times (\textsf{price}_t - \textsf{price}_{t-1}))_i$
    
    The optimality criterion is to maximize this empirical Sharpe Ratio. Transforming this into an optimization problem, the alpha that maximizes the objective function $\mathcal{L}(\alpha)$ is: 
    $$\Hat{\alpha} = \frac{\Sigma\textsuperscript{-1}\mu}{\sqrt{\mu^\text{T} \Sigma\textsuperscript{-1} \mu} } $$

    These optimal coefficients, upon examination, are seen to create a signal with the highest correlation to the asset's price variations, constrained by a low variance in this correlation. The constraint's trade-off is mecanically set to maximize the sharpe ratio. 
    
    \item This model can then be utilized as an unsupervised Machine Learning model for finance strategies: trained on a dataset composed of the price and exogenous variables of the last $\tau$ days $(t-\tau, \: ..., \: t-1)$, and used to create the present-time signal at $t$.

    Additional engineering techniques, mainly feature engineering of the exogenous variables, specific regularization techniques, and a corrective factor, are employed to develop and enhance the quantitative finance strategies based on this model.

    \item This model was tested to create a buy/sell strategy on a specific asset (\textsf{'IEF'}, a widely traded Exchange Traded Fund (ETF) that mirrors the performance of U.S. Treasury bonds with maturities of 1-3 years). The strategy yielded qualitative results, exhibiting good metrics in terms of risk-adjusted return: an effective Sharpe ratio of 1.2 when backtested over the period 2000-2023.

    However, a high turnover mitigates the quality of this specific strategy, indicating a need for further improvements.

\end{itemize}
\end{mdframed}

\newpage \:\vfill
\section{Problem Formulation}

Consider an asset whose price evolves according to the time series $(\textsf{price}_t)_t$. An investment strategy for this asset entails the creation of a time series $(\textsf{pos}_t)_t$, contingent on time, representing a position on this asset, i.e., the number of shares purchased of this asset multiplied by its price. Once a position $\textsf{pos}_t$ is determined, it is straightforward to compute the PnL: indeed, we have a asset value's variation term $(\textsf{pos}_t - \textsf{pos}_{t-1})$ which corresponds to the value variation of the positions and a cashflow term: $\textsf{price}_t\times(\frac{\textsf{pos}_{t-1}}{{\textsf{price}_{t-1}}}-\frac{\textsf{pos}_t}{\textsf{price}_t})$ which corresponds to encashing (resp. disbursing) the value difference of the positions sold (resp. bought) during the period. 

\noindent Hence, the sum of these two terms is: $\boxed{\textsf{PnL}_t = \textsf{pos}_{t-1} \frac{\textsf{price}_t - \textsf{price}_{t-1}}{\textsf{price}_{t-1}}}$

\smallbreak Subsequently, this time series $\textsf{pos}_t$ is primarily determined by constructing a principal signal in the form of a time series $\textsf{signal}_t$. Thus, in the general case, $\textsf{pos}_t = f_t(\textsf{price}_t, \textsf{signal}_t)$. It then becomes essential to ascertain both $f_t$ and $\textsf{signal}_t$ to optimize the PnL. This signal can be constructed in numerous ways, either technically or by utilizing fine economic relations between the asset price and certain exogenous variables. In our case, we will focus on the creation of a signal with a position-taking form as follow :
\begin{equation}
    \boxed{\textsf{pos}_t = \textsf{price}_t \times \textsf{signal}_t} \tag{H1}
\end{equation} which corresponds to the case where one decides to hold exactly $\textsf{signal}_t$ shares of the asset at any time\:$t$. Therefore we can deduce the relation between  $\textsf{PnL}_t$, $\textsf{price}_t$ and $\textsf{signal}_t$ : 
\begin{equation*}\boxed{\textsf{PnL}_t = \textsf{signal}_ {t-1}\times(\textsf{price}_t - \textsf{price}_{t-1})}
\label{linear relation equation}\end{equation*}

Now let's consider a set of \emph{stationary} and \emph{homoscedastic} variables denoted as $X_{1,t}, X_{2,t}, \ldots, X_{n,t}$ and create a signal as a linear combination of these variables. The sought-after signal is:
\begin{equation}\boxed{
    \textsf{signal}_t = \alpha_0 + \alpha_1 X_{1,t} + \alpha_2 X_{2,t} + \ldots + \alpha_n X_{n,t}}\label{linear signal hypothesis}\tag{H2}
\end{equation} 

\noindent That we will note with matrix notation:  $\textsf{signal}_t= \alpha^\text{T} X_t \text{ with: } X_{0,t} = 1\text{ for the intercept}$ \smallbreak 

\noindent Multiplying equation (\ref{linear signal hypothesis}) by $(\textsf{price}_t - \textsf{price}_{t-1})$ and then substituting relation between $\textsf{PnL}_t$, $\textsf{price}_t$ and $\textsf{signal}_t$, into yields: $$\boxed{\textsf{PnL}_t = \alpha^\text{T} \Tilde{X}_t}
\text{ where } \Tilde{X}_{i,t} = (\textsf{price}_t - \textsf{price}_{t-1}) X_{i,{t-1}}$$

\bigbreak
Now that a parametric expression for our PnL has been established, our goal is to define a metric for optimizing our PnL. The Sharpe Ratio, acts as a gauge for evaluating the risk-adjusted performance of a trading strategy. It is computed as the mean return of the strategy minus the risk-free rate (in practice in quant finance, and in this paper, the risk-free rate is set to $0$), divided by the standard deviation of the return, thus rendering a normalization for volatility. This ratio is usefull as it elucidates the return potential but also encapsulates the inherent risk, thereby offering a comprehensive measure of a strategy's efficacy and robustness.
\smallbreak The objective function we aim to maximize is thus defined as the Sharpe Ratio of the PnL, computed over the expectancy and standard deviation:
$$
\mathcal{L} := \frac{\mathbb{E}[\textsf{PnL}]}{{\sigma}[\textsf{PnL}]}
$$

\section{Mathematical Analysis and Optimization}

Using the notations $\mu^\text{T} = {\begin{bmatrix}
\:\overline{\Tilde{X}_{0,t}}, &
\overline{\Tilde{X}_{1,t}}, &
\ldots, &
\overline{\Tilde{X}_{n,t}}\:\end{bmatrix}}$ and $\Sigma = \begin{bmatrix}
\text{cov}(\Tilde{X}_{i,t}, \Tilde{X}_{j,t})_{i,j}
\end{bmatrix}$ for the empiric mean vector and covariance matrix of $\Tilde{X}_{t}$, we have: $$\overline{\textsf{PnL}} = \alpha^\text{T}\mu \text{   and   } \sigma_{emp.}{(\textsf{PnL})} = \sqrt{\alpha^\text{T} \Sigma \alpha}$$
\newpage
\noindent Thus, we can compute the objective function as the empiric sharpe ratio:
$$
\mathcal{L}(\alpha) =  \frac{\alpha^\text{T} \mu}{\sqrt{\alpha^\text{T} \Sigma \alpha}}
$$

\noindent For an invertible $\Sigma$, the closed-form solution for the $\alpha$ that maximize $\mathcal{L}$ is\footnote{
We can compute the gradient of $\mathcal{L}$ which is null when $\Sigma\alpha \propto \mu$: $$\nabla \mathcal{L}(\alpha) = \frac{\mu (\alpha^\text{T} \Sigma \alpha) - \Sigma\alpha}{(\alpha^\text{T} \Sigma \alpha)^{3/2}}$$ The coefficient of proportionnarity is free since $\mathcal{L}(|cst| \times \alpha) = \mathcal{L}(\alpha)$ we have choosen a positive one that gives: $\alpha^\text{T} \Sigma \alpha = 1$ for a unitary standard deviation of the PnL and a positive expectancy.\smallbreak}:
\begin{equation*}\boxed{
\Hat{\alpha} = \frac{\Sigma\textsuperscript{-1}\mu}{\sqrt{\mu^\text{T} \Sigma\textsuperscript{-1} \mu}}  
}\label{Optimal beta formula}\end{equation*}\nopagebreak\noindent Finally the Optimal Linear Signal extracted from the exogenous variables can be computed by: $\Hat{\alpha}^\text{T}X_t $


\begin{mdframed}
\begin{center}
    \bfseries How to Interpret These Results?
\end{center}

\noindent A 'good signal' is defined as a time series $\textsf{signal}_t$ whose 1-day lagged series $\textsf{signal}_{t-1}$ exhibits a high correlation with the asset's price variation $(\textsf{price}_t - \textsf{price}_{t-1})$. The methodology we present aims to construct a linear signal that maximally captures the most of this correlation using $n$ exogenous variables, under the constraint of a low variation of this correlation.

In effect, $\mu_i$ are estimator of $\mathbb{E}[(\textsf{price}_t - \textsf{price}_{t-1}) \times X_{i,t}]$, and actually represents the empirical covariance between the series $(X_{i,t-1})_t$ and the price variations $(\textsf{price}_t - \textsf{price}_{t-1})$. The corrective term involving $\Sigma\textsuperscript{-1}$ is introduced to reduce the variance of the PnL generated by this method. 

The trade-off is mecanically set to maximize the sharpe ratio. 
\end{mdframed}

\paragraph{Remarks:} \begin{itemize}
    \item The presence of non-stationarity in the $X_t$ variable introduces bias, predominantly captured by the intercept term $\alpha_0$. Essentially, the mean contribution of each variable influences the intercept, potentially assigning it an irrelevant value that fails to capture information effectively. 
    \item Heteroscedasticity can introduce bias in the model. This phenomenon occurs because variables with larger amplitude variations are perceived by the model as being more significant compared to others. Such disproportionate emphasis on certain variables due to their variance can skew the model's performance, leading to biased outcomes. 
\end{itemize}
Then, a good practice is to standardize the exogenous variables $X_t$ before using them in the model, it forces them into stationnarity and homoscedasticity, at the conditio, that the mean and variance does not change too fast -- i.e. the magnitude of changes is low on a timescale of 'training size'.

\subsection{Achieving beta neutrality}

A limitation of this optimizatio problem is that it is not inherently beta neutral. The signal might correlate with the asset's price or, by extension, the market. However, at the expense of some performance, it is feasible to introduce a constraint to decorrelate the signal from the asset's price.
\smallbreak
Incorporating the constraint $\alpha^{\text{T}}\beta = 0$ into the optimization problem alters the vector $\mu$ to $$\Tilde{\mu} = \mu - \frac{\mu^{\text{T}}\Sigma\textsuperscript{-1}\beta}{\beta^{\text{T}}\Sigma\textsuperscript{-1}\beta}\beta$$ and then use this altered $\mu$ in the optimal beta formula given in section \ref{Optimal beta formula}. 
\smallbreak 
By setting $\beta = \overline{\textsf{price}_t \times X_t }$, the average of the product of $X_t$ and $\textsf{price}_t$, we generate a signal uncorrelated with the asset's price, achieving a form of beta neutrality. We can also choose any $y_t$ and $\beta = \overline{y_t \times X_t }$ in order to get a signal uncorrelated to $y_t$.

\subsection{Regularization techniques}

\paragraph{L1 regularization:} When confronting potential redundancy among exogenous variables, L1 regularization emerges as a viable solution. It modifies the objective function to: $$\Tilde{\mathcal{L}}(\alpha) = \mathcal{L}(\alpha) - \lambda_1 |\alpha|$$ thereby promoting sparsity in the parameter vector $\alpha$. No closed-form solution is available in this case, necessitating numerical optimization to find: $ \Hat{\alpha} = \text{argmax}\:\Tilde{\mathcal{L}}$. To maintain parameter comparability and interpretability an alternative formulation can be considered: $$
\Tilde{\mathcal{L}}(\alpha) = \mathcal{L}(\alpha) -(\lambda_1\times\text{max}\:\mathcal{L})\times|\alpha|
$$

\paragraph{L2 regularization:} For non-invertible covariance matrix $\Sigma$, L2 regularization can be utilized to ensure invertibility by adding a scaled identity matrix, resulting in: $$\Tilde{\Sigma} = \Sigma + \lambda_2 Id$$ However, to preserve parameter comparability and interpretability, an alternative transformation can be employed: $$\Tilde{\Sigma} = \frac{(\Sigma + \lambda_2 \frac{||\Sigma||}{n} Id)}{(1+\lambda_2)}$$
\smallbreak



\paragraph{PCA regularization:} 

Another approach involves the use of PCA (Principal Component Analysis) before calculating $\mu$ and $\Sigma$, in order to select only the top $k$ principal components of $\Tilde{X}_t$. Caution is required: the PCA should be performed on $\Tilde{X}_t$ rather than $X_t$. This is because the valuable information resides in the covariance of the transformed variables $\Tilde{X}_{i,t} = (\textsf{price}_t - \textsf{price}_{t-1}) X_{i,{t-1}}$, not in the covariance of the original variables. Therefore, conducting PCA within the OLS model is preferable rather than prior to it. 

\smallbreak 
Denoting $\mu_i$ and $\sigma_{i}$, the empiric mean and standard deviation of the $i^{th}$ principal component. Since the principals components are uncorrellated, $\Sigma$ is diagonal and $\Hat{\alpha}$ is computed with : $$\Hat{\alpha_i} = \frac{\mu_i}{\sigma_{i}^2}/\sqrt{\sum_{j=1}^k\frac{\mu_j^2}{\sigma_{j}^2}}$$

\noindent The signal regularized is: $\Hat{\alpha}^\text{T} \Pi X_t$ with $\Pi$ the $(k\times n+1)$ matrix of the projector onto the top $k$ principal components of $\Tilde{X}_t$. Note that $\Pi$ is the projector onto the top $k$ principal components of $\Tilde{X}_t$, not of $X_t$.

\paragraph{Statistical Significance Regularization:} For each alpha coefficient, we compute the p-value of $\alpha_i$ and retain only those coefficients $\hat{\alpha}_i$ that meet the criterion:
$$\mathbb{P}[{\alpha}_i = 0 | \hat{\alpha}_i] < p_{\text{threshold}}$$

The p-value is computed assuming that $\Tilde{X_t}$ adheres to a Gaussian distribution. Under null hypothesis subsequent calculations\footnote{The demonstration is straightforward in the case where $\Sigma$ is diagonal because we have : $\sqrt{\tau}{\hat{\alpha}_i}\times{\hat{\alpha}^\text{T}\mu} = \sqrt{\tau}\frac{\mu_i}{\sigma_i}$.} reveal that $\sqrt{\tau}{\hat{\alpha}_i}\times{\hat{\alpha}^\text{T}\mu}$ follows a Student's t-distribution with $\tau - 1$ degrees of freedom, where $\tau$ denotes the length of the time series. This understanding enables the computation of the aforementioned p-value.

        
\begin{mdframed}
\begin{center}
    \bfseries Practical use of the ML model
\end{center}

\noindent In practical terms, this model is applicable in the development of a trading strategy. This process entails the training of the model over a span of $\tau$ days, utilizing datasets $(X_{t-\tau}, ..., X_{t-1})$ and $(\textsf{price}_{t-\tau}, ..., \textsf{price}_{t-1})$. From this training dataset, $\Hat{\alpha}$ is deduced through the prescribed methodology. Then, a signal for the subsequent day, $t$, is formulated as $\hat{\alpha}^\text{T} X_t$. 

This results in a machine learning model able to at generate linear trading signals from exogenous variables. The model is parametric, encompassing parameters such as the training period $\tau$, and a regularization parameter for each regularization technique. 
\end{mdframed} 

\paragraph{Remark on overfitting:} With at most an order of magnitude of $5000$ data points, the available dataset is quite limited, inherently posing a constant risk of overfitting. The key to the success of such a machine learning model lies in its ability to effectively implement regularization. This is the rationale behind proposing a variety of complementary regularization methods in the previous section. Beyond feature engineering, a significant portion of the empirical work presented in the next part has been dedicated to constructing and optimizing relevant regularization techniques to mitigate this issue. Still, this problem should be kept in mind as a important risk using this model for a real trading strategy.

\section{Empirical Application to a Trading Strategy}
 
The strategies empirically tested in this document is a day-to-day strategies. We consider an asset with an opening price denoted by $\textsf{price}_t$, reflecting the Open price of the asset in the market. A signal $\textsf{signal}_t$ is generated using the proposed method. At the market opening we buy\footnote{Note: The strategy operates under the assumption that: \emph{(i)} the asset can be acquired at its opening price on day $t$, \emph{(ii)} without any transaction fees, and that \emph{(iii)} the action of the strategy will not have any effects on the market price. }, $\textsf{signal}_t$ shares of the asset, under the condition that the signal is sufficient, establishing a position of: $$\textsf{pos}_t = \textsf{signal}_t \times \textsf{price}_t \text{ if: } Z_{score}(\textsf{signal}_t)>1 \text{ else: }0$$ 

\begin{mdframed}
    \noindent\textbf{Traded Products:} \textsf{'IEF'}, a widely traded Exchange Traded Fund (ETF) that mirrors the performance of U.S. Treasury bonds of maturities: 1-3 years. \smallbreak\noindent\textbf{Exogenous variables:} Prices of U.S. treasury bonds of differing maturities and various macros, \emph{that have been feature-engineered.} \smallbreak\noindent\textbf{Data:} Open source data from Yahoo finance.
\end{mdframed} 

\medbreak
The graph below corroborates the signal's efficacy: intervals of augmented signals (illustrated in red) are congruent with noticeable ascents in the PnL.
\begin{center} \includegraphics[width=\textwidth]{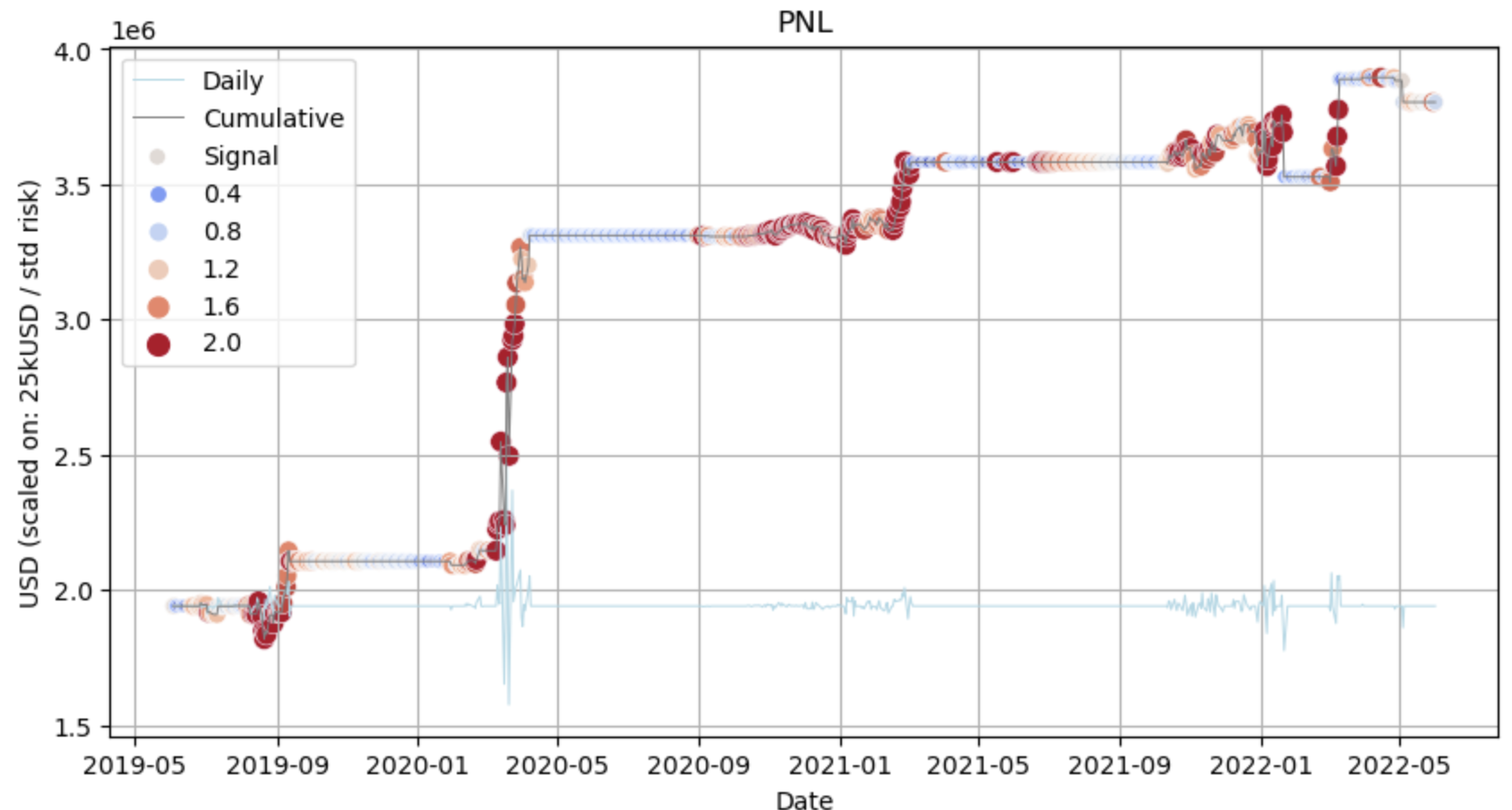}
Quantitative Metrics: Sharpe Ratio: 1.25 / effective\footnote{Effective metrics are computed over the days where and effective trading has been made i.e. at days $t$ where $\textsf{pos}_t \neq 0$}: 2.1 ;\linebreak Turnover: 45.9\%; Bips: 18.9 
\end{center} 

\smallbreak  However, it is observed that over a longer period, we are subject to fluctuations: at times, the signal misinterprets the direction. The primary reason is that the signal can detect that the period is conducive to generating PnL, but it remains highly uncertain about which direction to take. Consequently, during each of these periods, the signal can choose the wrong direction, leading to frequent monetary losses in the strategy, which adversely affects the metrics. As a result, we obtain rather mediocre outcomes, as shown in the following figure.

\begin{center} \includegraphics[width=\textwidth]{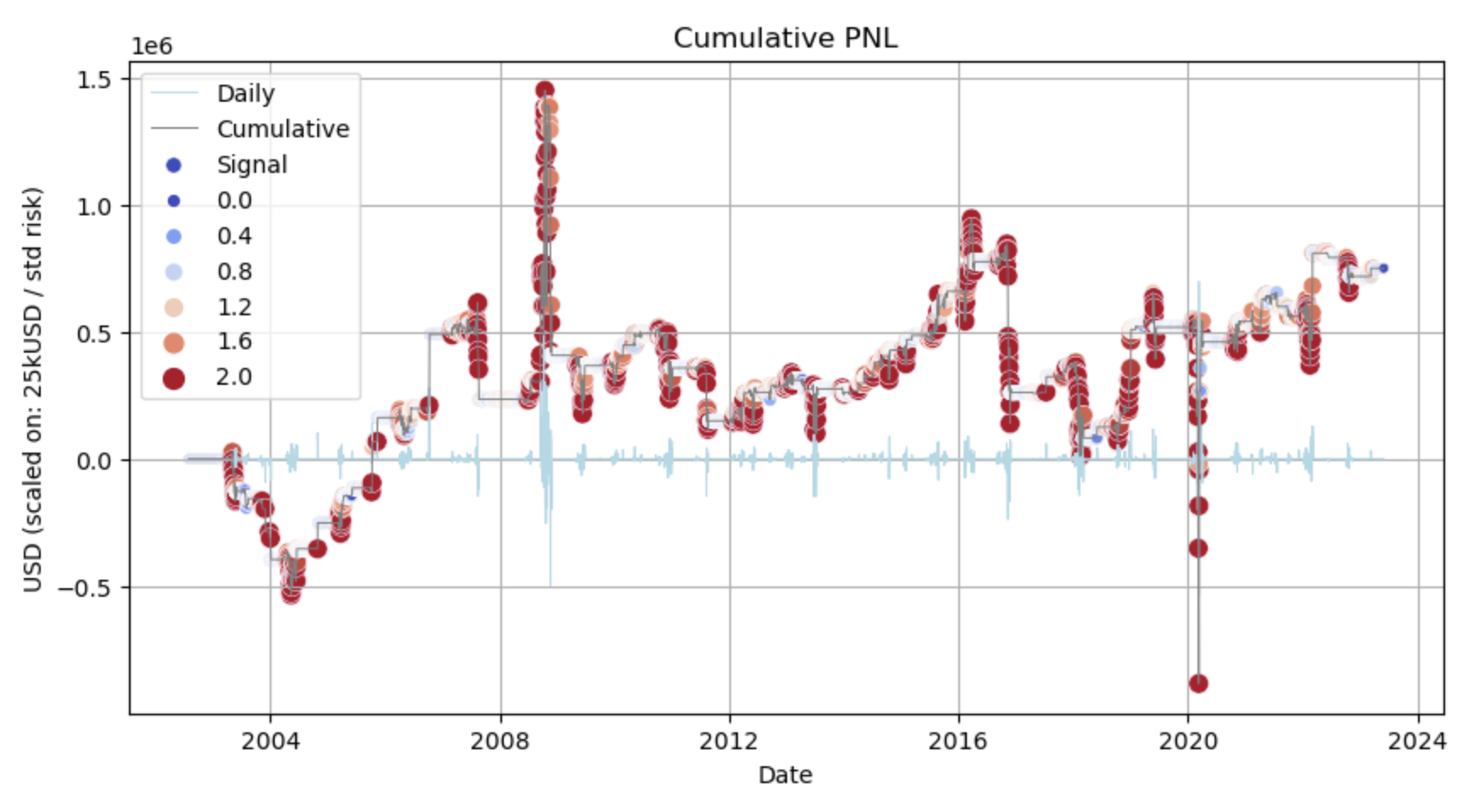}
Quantitative Metrics: Sharpe Ratio: 0.09 / effective: 0.46 ;
\end{center} 
\bigbreak

Nevertheless, as depicted in the graph below, two corrective techniques have been identified that drastically improve the results: 
\begin{itemize}
\item the introduction of a corrective factor, represented as `$\textsf{signal}_t = \text{sg}(\textsf{PnL}_t) * \Hat{\alpha}^\text{T}X_t $' (if the uncorrected model would have lost money on the previous day, the signal is reversed.) and, 
\item the intensive application of statistical significance regularization, 
\end{itemize}
    
\begin{center} \includegraphics[width=\textwidth]{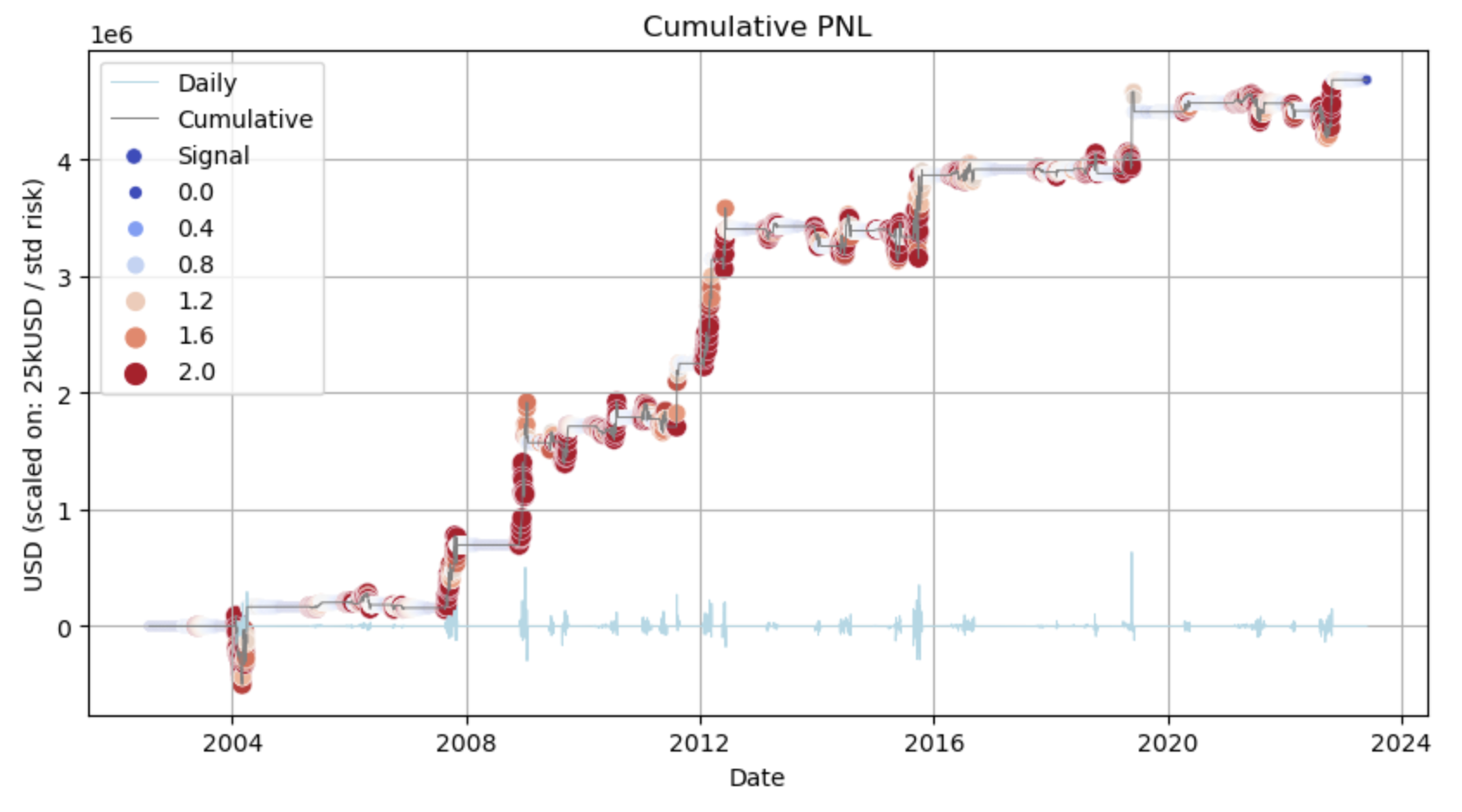}
\bfseries Quantitative Metrics: Sharpe Ratio: 0.55 / effective: 1.19 ;\linebreak Turnover: 37.3\%  / effective: 184.1\% ; Bips: 11.9 / effective: 51.1
\end{center} 

To verify the functionality of our strategy beyond a specific set of parameters and to identify the optimal range of parameters yielding the best possible results, we conducted a comprehensive analysis by calculating the Sharpe Ratio of the PnL for various training sizes. This exercise is crucial to ensure that the strategy was not just effective for a particular set of parameters (overfitting) but also adaptable and robust across different scenarios.

\begin{center} \includegraphics[width=\textwidth]{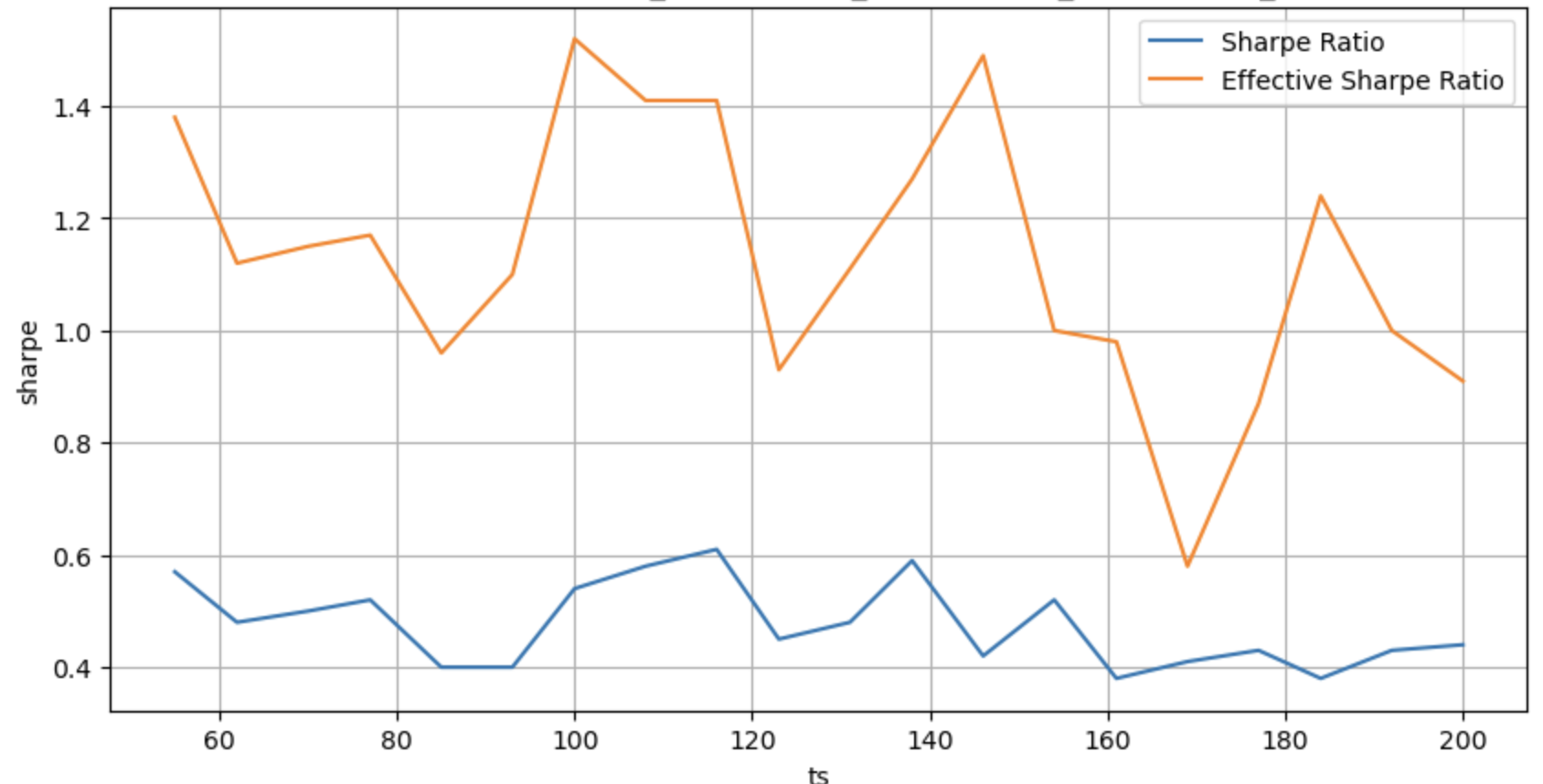}
\bfseries Sharpe and effective sharpe ratio for different training sizes
\end{center} 



\paragraph{Caveat:} this strategy is good in terms of Sharpe ratio. Yet, when we look at the effective turnover, we realize that these strategies perform poorly from this indicator's perspective (effective turnover is more than 150\%.) Indeed, when plotting the PnL curve versus the curve of positions taken, it becomes apparent that the strategy requires taking large positions to generate significant PnL.

\begin{center} \includegraphics[width=\textwidth]{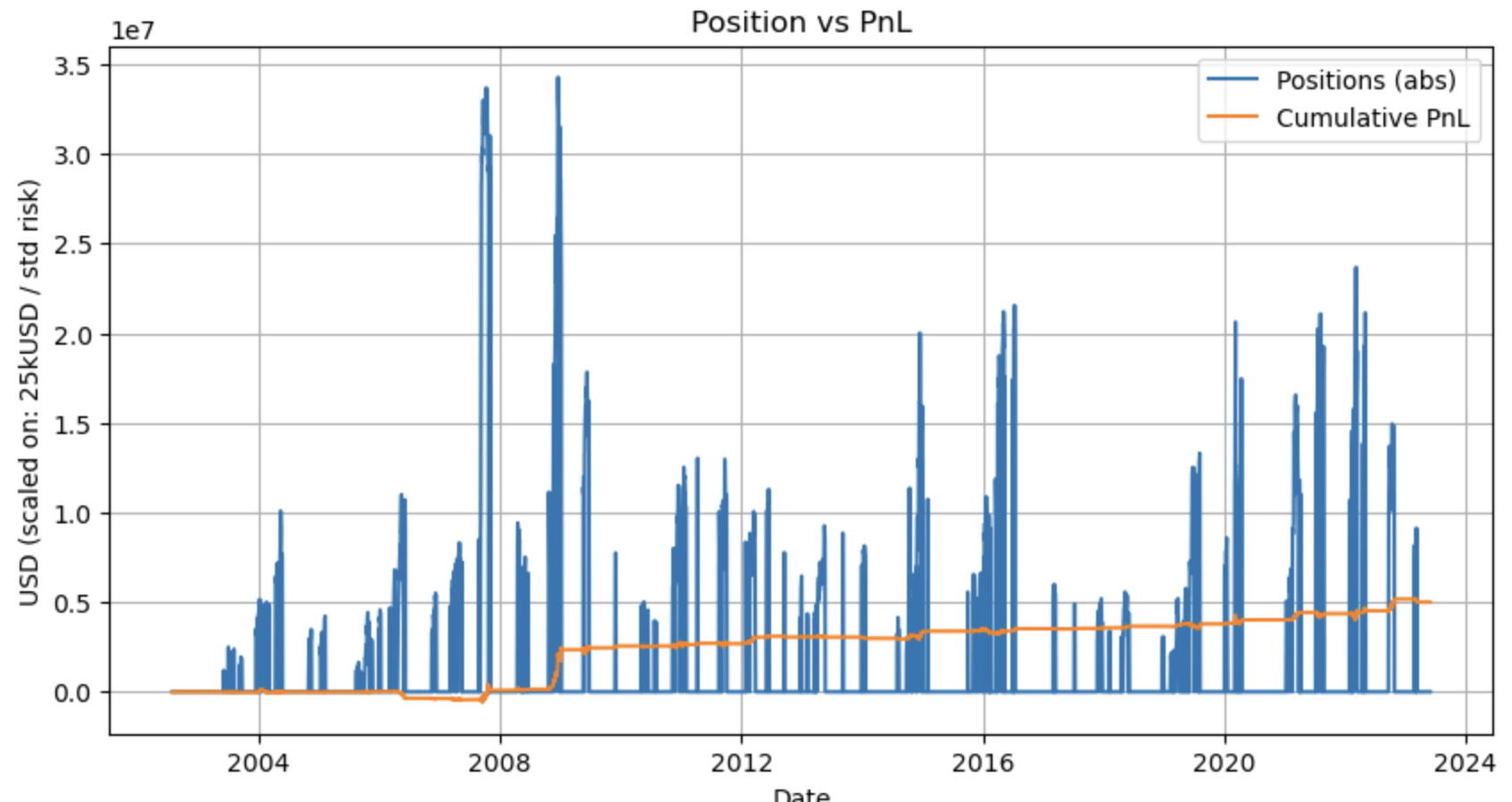}
\end{center}

In this specific scenario, to generate a cumulative PnL of 5 million USD, there are brief periods where an investment of up to 35 million USD is required. This observation underscores the strategy's demand for significant capital deployment at sporadic intervals to achieve the desired returns. 

Nevertheless, given the high degree of certainty in these strategies (a Sharpe ratio of 1 implies that money is lost in less than 7\% of cases), it becomes feasible to obtain such investments through leverage. This approach allows for maximizing potential returns while managing the risk associated with significant capital deployment. Thus, the described strategy either demands substantial sporadic investments, via leverage effect, or it needs to be refined through other quantitative finance methods to reduce this turnover. 

\section*{Conclusion}

This study delineates a comprehensive framework for systematically constructing and optimizing trading strategies, anchored in a solid mathematical approach. The core innovation resides in the formulation of a signal linearly derived from selected exogenous variables. By integrating a linear combination of these variables into a strategic position-taking model, we have demonstrated a method that effectively constructs a signal responsive to market dynamics and predictive of future trends.

The methodology involved formulating the trading signal as a linear combination of a set of exogenous variables, which was subsequently employed to ascertain the position in the asset. The consequent PnL computation, based on the asset's price variations and the held positions over time, underscores the robustness of the proposed framework. This was further validated through empirical application, using publicly accessible market data, to demonstrate the efficacy of the model in constructing a durable trading strategy. Looking forward, the study opens avenues for further technological development, which include:

\begin{itemize}
    \item \textbf{Implement a New Regularization Aimed at Limiting Position Size:} This approach could address the high-turnover issue observed in the empirical tests. By integrating a regularization mechanism that constraints the size of positions taken, the strategy can be fine-tuned to balance PnL generation against the risk of excessive capital allocation. It would result in a sharpe ratio / turnover trade-off. Such regularization would ensure that the strategy remains robust and efficient, without necessitating disproportionately large investments.

    \item \textbf{Adapting the Model for Long/Short Strategies:} Beyond the projection method described in subsection 2.1 to achieve beta neutrality, it is feasible to adapt the model for executing long/short strategies. This requires considering a multi-layer model, where an initial signal is generated for each asset. Subsequently, one must weight the signals of each asset to maintain a balanced position situation. An analogous algorithm can be employed to optimally select this weighting.


    \item \textbf{Generalized Time Steps:} In the 'How to Interpret These Results' section, we have shown that this methods constructs a time series $s_t$ such that its 1-day lagged series $s_{t-1}$ is strongly correlated with the asset price variations $(\textsf{price}_t - \textsf{price}_{t-1})$. This concept can be extended to $t_s$-days lagged time series $s_{t-t_s}$, by examining their correlation with $(\textsf{price}_t - \textsf{price}_{t-t_s})$. The optimal $t_s$ intervals could be determined using methods that study seasonality, such as Fourier analysis.
    
    \item \textbf{Generalized Linear Signal:} Expanding the signal model to a more generalized form: $\textsf{signal}_t= f_{\text{act.}}(\alpha^\text{T} X_t)$, lead to a new expression for PnL. The samed process can be applied by optimizing the objective function $\mathcal{L}(\alpha)$ accordingly. 
    
    Some activation function $f_{\text{act.}}$ such as a logit activation function inspired by Logistic Regression to create a signal that can take values in [0,1], and can be used as a mitigateur of other signals. This generalization can be explored with various activation functions used as hyperparameters.

    \item \textbf{Continuous Hyperparameter Selection:} In order to mitigate the risk of an overfitting signal on a hyperparameter set, we can implement a dynamic system for adjusting hyperparameters every 'training-size' days, based on the hyperparameters that maximized metrics in the previous period. Note that this bruteforce method will significantly increase the computational cost.
    
    \item \textbf{Enhance Corrective Terms with Command \& Control Theory:} A primary limitation of this model is its relatively low reactivity, as it updates only every 'training-size' days. An approach grounded in command \& control theory could significantly enhance the model's responsiveness, enabling real-time adjustments. The 'corrective factor' introduced in the 'Empirical Application to a Trading Strategy' section represents a rudimentary form of this adaptive correction. Investigating the application of command and control theory, especially the use of Kalman filters, to refine the model's corrective terms emerges as a promising avenue for development. Such an advancement would not only add dynamism to the model but also improve its accuracy and adaptability in response to evolving market conditions.
    
    \item \textbf{Stacking Linear Signal:} Inspired by the concept of stacking regressors used in machine learning, we could create a new, stacked signal by utilizing the Optimal Linear Signal with already efficient signals used as Exogenous Variables, potentially enhancing the signals' strength. 
    
    In particular, by utilizing the 'Achieving Beta Neutrality' section, one can derive a beta-neutral signal from general signals.
    
    \item \textbf{Boosting Method:} The subsection 'Achieving Beta Neutrality' develops a method to generate a signal uncorrelated to a certain time series. If a signal is generated from n variables and a PnL is deduced from it, it is entirely possible to generate a new residual signal that is uncorrelated to this previous PnL. The sum of these two signals should have been able to capture more relevant information from the variables, on the condition that some non-linearity has been introduced somewhere, and sufficient regularization avoid overfitting.

    On the same principle as Boosting in ML, by iterating each time on the residual signal (orthogonal to the previous one), it is conceivable to construct very powerful signals.

\end{itemize}

In essence, this study not only lays a foundation for traders and financial analysts seeking to augment their strategies through quantitative methods but also sets the stage for continuous innovation and advancement in systematic trading strategies.

\end{document}